\documentclass[journal=jpccck,manuscript=article,layout=traditional]{achemso}

\usepackage[latin9]{inputenc}
\usepackage{color}
\usepackage{amsmath}
\usepackage{graphicx}
\usepackage{amsfonts}

\author{Kriszti\'an Szab\'o}
\affiliation{Department of Theoretical Physics, University of Debrecen, P.O. Box
400, H-4002 Debrecen, Hungary}

\author{Csaba F\'abri}
\affiliation{Department of Theoretical Physics, University of Debrecen, P.O. Box
400, H-4002 Debrecen, Hungary}

\author{G\'abor J. Hal\'asz}
\affiliation{Department of Information Technology, University of Debrecen, P.O.
Box 400, H-4002 Debrecen, Hungary}

\author{\'Agnes Vib\'ok}
\affiliation{Department of Theoretical Physics, University of Debrecen, P.O. Box
400, H-4002 Debrecen, Hungary}
\altaffiliation{ELI-ALPS, ELI-HU Non-Profit Ltd, H-6720 Szeged, Dugonics t\'er 13, Hungary}
\email{vibok@phys.unideb.hu}

\title{Indirect probing of light-induced nonadiabatic dynamics in lossy nanocavities}

\begin{document}

\begin{tocentry}
\includegraphics[scale=1.0]{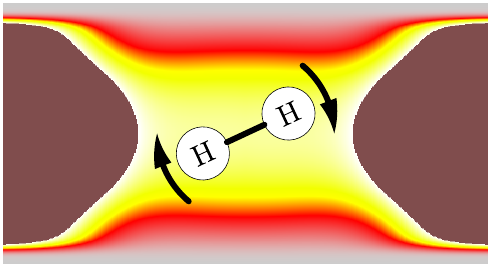}
\end{tocentry}

\begin{abstract}
Light-induced nonadiabatic effects can arise from the interaction of a molecule with the quantized electromagnetic field of a Fabry--P\'erot or plasmonic nanocavity. In this context, the quantized radiation field mixes the vibrational, rotational, and electronic degrees of freedom. In this work, we investigate the photodissociation dynamics of a rotating hydrogen molecule within a lossy plasmonic nanocavity. We highlight that, due to significant cavity loss, the dynamics are governed by an infinite number of light-induced conical intersections.
\textcolor{red}{}{We also examine the dissociation dynamics of fixed-in-space molecules by neglecting  rotation, employing both the Lindblad master and non-Hermitian lossy Schr\"odinger equations. Additionally, we incorporate the effects of rotation within the parameter range of perfect agreement using the non-Hermitian lossy Schr\"odinger method.}
Furthermore, we show that in the absence of photon losses, there is a close correspondence between the classical Floquet description and the radiation field model.
\end{abstract}

\section{Introduction}

When molecules are placed into optical or plasmonic nanocavities, hybrid light-matter states,
known as polaritons, are formed, exhibiting both
excitonic and photonic characteristics. 
The electric dipole moment of
the molecular system and the quantized radiation mode of the cavity
can resonantly couple, which enables the manipulation and control of various
physical and chemical properties of molecules. Numerous experimental
studies \cite{12HuScGe,16ChNiBe,16Ebbesen,16ThGeSh,16VeGeCh,16ZhChWa,17ChThAk,18GrTo,19DaWeKr,19OjChDe,19RoShEr,19VeThNa,24JaReSi}
and theoretical investigations \cite{15GaGaFe,16GaGaFe,16KoBeMu,17FlApRu,17HeSp_2,18FeGaGa,18PiScCh,18RuTaFl,18SzHaCs_2,18Vendrell,19ReSoGe,19TrSa,21HaAhBe,21TrSa,22RiHaRo,22MaAkBe,23Fabri,23ScKo_2,23ScKo,23ScSiRu,23Szidarovszky,23Szidarovszky_2,24FiRi,24SoGr, 24BaReHo,25Szidarovszky,25FaHaHo}
have focused on molecular polaritons since the first pioneering experimental work 
reported by Ebbesen and coworkers.\cite{12HuScGe,16ZhChWa}
Among the key findings, the quantized
radiation field has been shown to amplify \cite{24SaFeGa} or suppress
\cite{16GaGaFe} well-known physical mechanisms and even induce novel
effects. It can significantly alter the rate of spontaneous emission
\cite{21Wang}, energy transfer \cite{18DuMaRi,21LiNiSu}, and charge
transfer processes \cite{19SeNi,20MaKrHu,21WePuSc}. Moreover, strong
light-matter coupling has the potential to modify chemical landscapes
and reactions \cite{16KoBeMu,19CaRiZh,19GaClGa,22LiCuSu,23FrCo},
influencing processes such as photochemical reactivity \cite{20FeFrSc,20FrCoPe},
photoisomerization \cite{18FrGrCo,20FrGrPe}, photodissociation \cite{20DaKo_2,21ToFe},
photoionization \cite{23FaHaCe}, and photoassociation \cite{24CeFe}.
Additionally, cavity-molecule coupling can induce nonadiabatic effects in molecules 
that are absent under field-free conditions \cite{17CsHaCe_2,18SzHaCs_2,19CsKoHa,20GuMu,20GuMu_2,20SzHaVi,21FaMaHu,21SzBaHa,22BaUmFa,22CsVeHa,22FaHaCe,22FaHaVi,22FiSa,23HaXiHu,24FaCsHa_2,25WaWaBe}. 
In these cases, the quantized radiation field can resonantly couple
molecular electronic, vibrational, and rotational degrees of freedom (DOFs),
leading to the formation of light-induced avoided crossings (LIACs)
or light-induced conical intersections (LICIs) \textcolor{red}{}{\cite{08MoSiCe,11SiMoCe,15HaViCe,16NaWaPr}}. 
In polyatomic molecules, a
sufficient number of vibrational DOFs are always present
to establish a two-dimensional branching space (BS) which is indispensable to form LICIs. 
In case of diatomic molecules, the rotational angle between
the molecular axis and the polarization direction of the cavity
field can serve as a dynamical variable to span the BS. 
A vast number of experimental and theoretical works have demonstrated
that the LICI has a remarkable impact on absorption spectra
(e.g., intensity borrowing) \cite{18SzHaCs_2,21FaHaCe,21FaHaCe_2,23FaHaCe},
topological properties (e.g., Berry phase) \cite{22BaUmFa,21FaMaHu},
and quantum dynamics of molecules \cite{19CsKoHa,22CsVeHa,22FaHaVi}. 
However, light-induced nonadiabatic dissociative dynamics in a lossy plasmonic
nanocavity has not been investigated yet. This paper aims to fill this gap.

To effectively manipulate and control molecular properties, it is
essential to reach the strong coupling regime where the rate of energy
exchange between the cavity photons and the molecule exceeds both
the photon leakage rate and system dephasing.
Strong coupling
can be achieved either by allowing a large number of molecules to
interact with the electromagnetic mode \cite{20UlVe,18FeGaGa}, or,
in case of a single molecule, by using a subwavelength plasmonic
nanocavity \cite{16ChNiBe}. However, the latter often exhibits significant
losses, which must be properly accounted for in the numerical treatment
\cite{20UlVe,20FeFrSc,21ToFe}.

Typically, photon losses are modeled by coupling the cavity-molecule
system to a dissipative Markovian environment, leading to the Lindblad
master equation approach \cite{20SiPiGa,20DaKo_2,20Manzano,21ToFe,22FaHaVi}.
In this framework, the time evolution of the system is described by
propagating the density matrix according to the appropriate Lindblad
master equation. Another common method is based on the non-Hermitian time-dependent
Schr\"odinger equation (TDSE), although this approach is not universally
applicable. The two methods are rigorously equivalent only when the
incoherent decay terms are absent from the dynamics. In such cases,
the quantum system is confined to a specific excitation manifold (e.g.,
the singly-excited subspace, consisting of a ground-state molecule
with one photon and an excited-state molecule with zero photons). Here,
the dissipative effects are incorporated into the non-Hermitian Hamiltonian,
which results in the loss of norm in the nuclear wave packet during
TDSE time propagation. In a previous study \cite{24FaCsHa}, we demonstrated
the additional conditions required for the Lindblad and non-Hermitian TDSE approaches to yield
comparable results.

In this study, we investigate the cavity-induced nonadiabatic photodissociation
dynamics of the hydrogen (H$_2$) molecule.
We perform both \textcolor{red}{}{fixed-in-space} and two-dimensional (2D) calculations where the molecular rotation is treated either as a fixed parameter ($\theta=0$, with $\theta$ being the angle between the polarization direction of the cavity electric field and the molecular axis) or as a dynamical variable in the 2D case. 
%
%
Our primary focus is short-time dynamics, as conical intersections
play a significant role on this timescale. It is worth noting that
while the photodissociation process of $\mathrm{H_{2}}$ has been
explored within the molecular plasmonic framework \cite{21ToFe},
previous studies have typically  neglected the effect of rotational DOFs
on nonadiabatic dynamics. 

The objectives of this paper are twofold. First, we aim to study
the cavity-induced nonadiabatic dynamics in a lossy plasmonic nanocavity. 
Such a study must be performed in 2D where LICIs appear. This will allow
us to reveal the difference between the impact of the cavity-induced
avoided crossings (LIACs) and of light-induced conical intersections (LICIs). 
Second, we also explore the dynamics
in the absence of photon losses, which allows us to demonstrate the
close similarity between the classical Floquet description and the
radiation field model \cite{24FaCsHa_2}.


\section{Theory and Computational Protocol}

A molecule coupled to a single cavity mode can be described by the Hamiltonian
\begin{align}
  \label{Eq:H-general}
  \hat{H} = \hat{T}_\textrm{M} + V_\textrm{M} + \hbar\omega_\textrm{c} \hat{a}^\dag \hat{a} - \vec{E} \vec{{\mu}} (\hat{a}^\dag + \hat{a}),
\end{align}
where $\hat{T}_\textrm{M}$ denotes the kinetic energy operator of the molecule (including vibration and rotation as well), $V_\textrm{M}$ is the potential energy surface (PES), $\omega_\textrm{c}$ is the cavity angular frequency, while $\hat{a}^\dag$ and $\hat{a}$ are the creation and annihilation operators of the cavity mode, respectively. 
The electric field strength is denoted by $\vec{E}$ and $\vec{\mu}$ stands for the transition dipole moment (TDM) of the molecule. We consider two electronic states of the hydrogen molecule, the ground electronic state $|\textrm{X}\rangle$ ($\textrm{X} \, ^{1}\Sigma_{g}^{+}$) and the excited electronic state $|\textrm{B}\rangle$ ($\textrm{B} \, ^{1}\Sigma_{u}^{+}$), both of which are singlet states. The PESs and TDM have been taken from Refs. \citenum{03WoSt}, \citenum{10Pachucki} and \citenum{21SiZiPa}. We treat the light-matter interaction within the electric dipole approximation.



\begin{figure}
\begin{centering}
\includegraphics[width=0.44\textwidth]{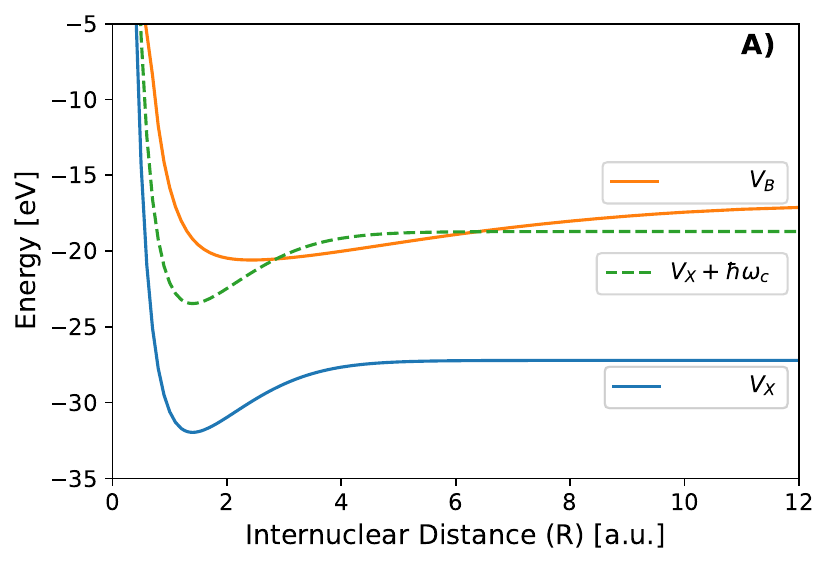}\\
\includegraphics[width=0.44\textwidth]{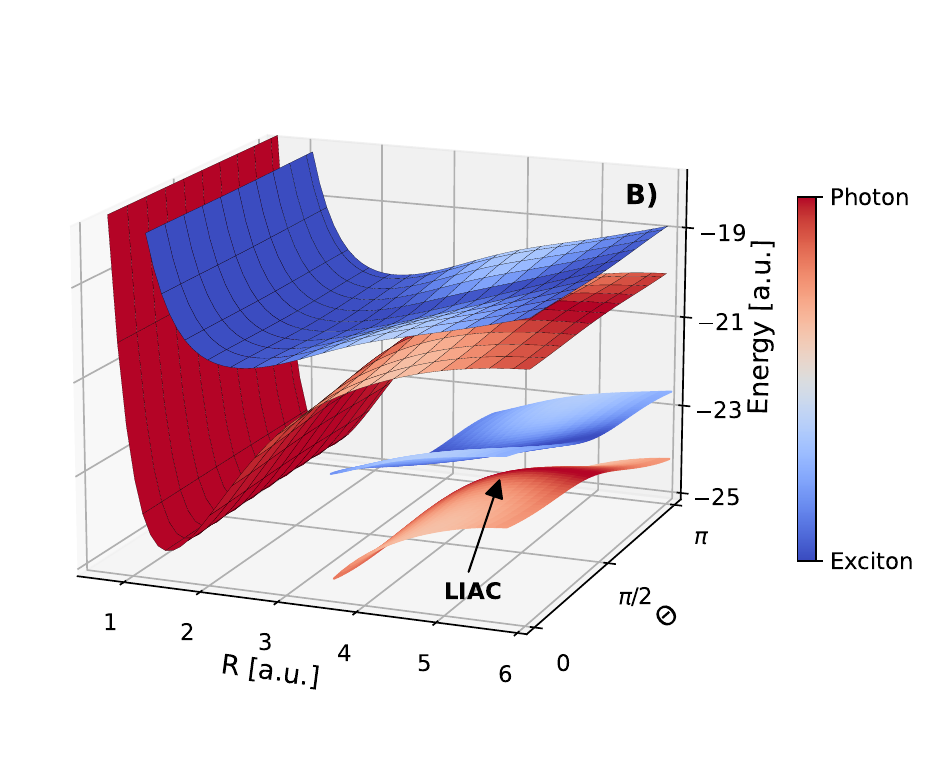}\\
\includegraphics[width=0.44\textwidth]{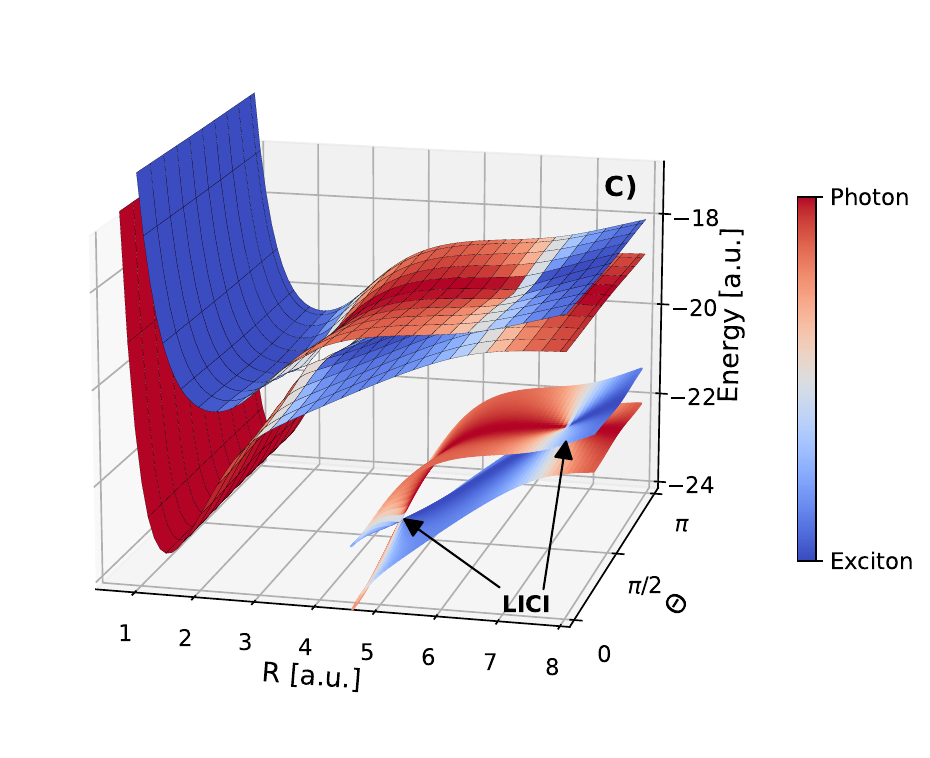}
\par\end{centering}
\caption{\textcolor{red}{}{Potential energies of the $\mathrm{H_{2}}$ molecule in a cavity.}
Panel A: Ground electronic state (denoted by $V_\textrm{X}$),
first singlet excited electronic state (denoted by $V_\textrm{B}$) and potential
energy curve corresponding to the molecule in ground electronic state
and one photon in the cavity. The cavity central photon energy is
$\hbar\omega_\textrm{c}=7.4\,\textrm{eV}$.
Panel B: Two-dimensional polaritonic potential energy surfaces (PESs) along
the vibrational and rotational coordinates of the $\mathrm{H_{2}}$ molecule.
The light-induced avoided crossing (LIAC) is highlighted in the inset on the right. 
The character of the polaritonic potential energy surfaces is indicated by different
colors (see the legend on the right).
The cavity central photon energy and field strengths are $\hbar\omega_\textrm{c} = 7.4\,\textrm{eV}$ and $E=300\,\mathrm{mV \slash a_0}$ \textcolor{red}{}{$\approx 1.1\cdot10^{-2}$ a.u.}.
Panel C:
Similar to panel B, but with cavity central photon energy of $\hbar\omega_\textrm{c} = 8.5\,\textrm{eV}$.
The light-induced conical intersections (LICIs) are highlighted in the inset on the right.
}
%
\label{fig:1} 
\end{figure}

The polaritonic or dressed states can be calculated by diagonalizing the
Hamiltonian of Eq. \ref{Eq:H-general} without the kinetic energy term in the so-called singly-excited subspace which is spanned by the set $\{|\textrm{X},1\rangle , |\textrm{B},0\rangle\}$. We use the notation $|i,n\rangle = |i\rangle \otimes |n\rangle$ with $i=\textrm{X},\textrm{B}$ and $n=0,1,\dots$ for the combined states of the molecule and cavity, where $|n\rangle$ denotes the Fock state of the cavity mode. 
In this case, the potential energy matrix has two eigenvalues for each nuclear configuration $(R,\theta)$, which span two surfaces over the branching space: lower (LP) and upper polaritonic (UP) surfaces, shown in Fig. \ref{fig:1} (panel B).


The strong coupling regime (for a single molecule) can be reached using a plasmonic nanocavity which typically has a much shorter lifetime than the timescale of the molecular dynamics. Therefore, the dissipative nature of the cavity has to be taken into account. Under these conditions, the time evolution of the system is governed by the Lindblad master equation\textcolor{red}{}{\cite{20SiPiGa,20DaKo_2,20Manzano,21ToFe,24FaCsHa}}
\begin{align}
  \label{Eq:Lindblad-ME}
  \frac{\partial \hat{\rho}}{\partial t} = -\frac{\textrm{i}}{\hbar} \left[\hat{H}, \hat{\rho} \right] + \frac{1}{2} \sum_{i} \left( 2 \hat{C}_i \hat{\rho} \hat{C}_i^\dag - \hat{\rho} \hat{C}_i^\dag \hat{C}_i - \hat{C}_i^\dag \hat{C_i} \hat{\rho} \right)
\end{align}
where $\hat{\rho}$ is the density operator describing the system, $\hat{H}$ is the Hamiltonian and $\hat{C}_i$ denotes collapse operators. We assume that the molecular excitations have infinite lifetime, hence only the cavity is coupled to the environment with $\hat{C}_\textrm{c} = \sqrt{\kappa} \hat{a}$, describing photon leakage to the environment. \textcolor{red}{}{Here, the cavity decay rate $\kappa$ ranges from $0$ to $0.476$ eV (equivalent to a finite photon lifetime $\tau=\hbar/\kappa$, e.g. $\kappa=0.476$ eV $\rightarrow~\tau\approx 1.38$ fs) and $\hat{a}$ is the usual annihilation operator of the cavity mode.}
The Lindblad equation was solved within the QuTip framework \cite{12JoNaNo,13JoNaNo} \textcolor{red}{}{(fixed-in-space model)}, 
where we used the truncated Hilbert space of $\{ |\textrm{X},0\rangle, |\textrm{X},1\rangle, |\textrm{B},0\rangle, |\textrm{B},1\rangle \}$. The vibrational DOF (internuclear distance, denoted by $R$) was represented using a sine DVR (discrete variable representation) basis \cite{00LiCa}, from $R=0.5$ a.u. to $R=17.0$ a.u. with $N=200$ grid points, while the cavity mode was handled in the Fock representation.


Our simple case of a two-state hydrogen molecule with one cavity mode is already considered a challenging system. In special cases, the dynamics can be properly approximated with the time-dependent Schr{\"o}dinger equation (TDSE) by incorporating an imaginary term in the Hamiltonian of Eq. \eqref{Eq:H-general}. In our case the TDSE takes the form\textcolor{red}{}{\cite{20FeFrSc,20UlVe,24FaCsHa}} 
\begin{equation}
\textrm{i} \hbar \frac{\partial |\Psi\rangle}{\partial t} =
    \Bigl( \hat{H} - \textrm{i} \frac{\kappa}{2} \hat{N} \Bigr) |\Psi\rangle.
	\label{eq:Schrodinger}
\end{equation}
The loss term works similarly to a complex absorbing potential (CAP), the flux absorption is proportional to the physical loss rate $\kappa$ and the number of photons present in the cavity (given by the number operator $\hat{N}=\hat{a}^\dag \hat{a}$). 
Of course, this is an approximation of the true Markovian interaction. 
It is also shown in Ref. \citenum{95ViNi} that the Lindblad master equation without the term
$\kappa \hat{a} \hat{\rho} \hat{a}^\dag$
is equivalent to the TDSE of Eq. \eqref{eq:Schrodinger}.
The term $\kappa \hat{a} \hat{\rho} \hat{a}^\dag$ induces incoherent transitions $|\alpha (n+1)\rangle \rightarrow |\alpha n\rangle$ ($\alpha$ labels
molecular electronic states X and B).
The TDSE (see Eq. \eqref{eq:Schrodinger}) was solved with the multiconfiguration time-dependent Hartree (MCTDH) method \cite{90MeMaCe,00BeJaWo}.
As basis, we used sine DVR for the vibrational DOF from $R=0.5$ a.u. to $R=17.0$ a.u. with $N=500$ grid points, Legendre DVR for the rotational DOF with $N_\theta=100$ grid points and Hermite DVR to describe the cavity mode with $N_\textrm{c}=10$ grid points.


The dissociation of the H$_2$ molecule can lead to the reflection of the wave function at the edge of the grid, therefore we employ complex absorbing potentials (CAP) for each electronic state. The usual form of a CAP is \textcolor{red}{}{\cite{00BeJaWo}}
\begin{align}
  \label{Eq:CAP-standard}
  -\textrm{i} W(R) = -\textrm{i} \eta |R - R_\textrm{c}|^b \theta \left(R - R_\textrm{c} \right)
\end{align}
where $R$ is the coordinate of the dissociative DOF, $\theta (x) $ denotes the Heaviside step function, $\eta$ is a scalar, referred to as the strength of the CAP and $R_\textrm{c}$ is the grid point where the CAP is switched on. Our parameters are chosen as $\eta = 10^{-4}$, $R_\textrm{c} = 12$ a.u. and $b=4$.

The Lindblad equation  describes strictly trace-preserving dynamics, which means that the Hamiltonian used in Eq. \eqref{Eq:Lindblad-ME} must be Hermitian. Therefore, one can not use a CAP (see Eq. \eqref{Eq:CAP-standard}) in this case. We can overcome this issue with a series of collapse operators, each couples one grid point to a special basis function $|\mathcal{D} \rangle$, with coupling strength proportional to $W(R_i)$ of Eq. \eqref{Eq:CAP-standard} with $R_i$ being the coordinate of the given grid point. The absorbing potential \textcolor{red}{}{used in the Lindblad method}, equivalent to the CAP of Eq. \eqref{Eq:CAP-standard}, is given by the \textcolor{red}{}{series of} collapse operators \textcolor{red}{}{\cite{13JoNaNo}}
\begin{align}
  \label{Eq:CAP-Lindblad}
  C_\textrm{Abs} := \{ \sqrt{2 W(R_i)} | \mathcal{D} \rangle \langle R_i | \} ~~~~ \forall i, R_i > R_\textrm{c}
\end{align}
where $|R_i\rangle$ is the internuclear DVR basis function corresponding to the grid point $R_i$.


The main observables we are interested in are the populations ($P_{\textrm{A},i}$) and integrated dissociation probabilities ($P_{\textrm{D},i}$). 
The population of states in case of the TDSE can be calculated as a simple inner product of the wave function corresponding to the given electronic state at every time step,\textcolor{red}{}{\cite{94Sakurai}}
\begin{align}
  \label{Eq:PA_TDSE}
  P_{\textrm{A},i} (t)= \langle \Psi_i (t) | \Psi_i (t) \rangle
\end{align}
where $i = \textrm{X}, \textrm{B}$.
The integrated dissociation probability can be calculated through the so-called flux analysis method, which gives \textcolor{red}{}{\cite{00BeJaWo}}
\begin{align}
  \label{Eq:PD_TDSE}
  P_{\textrm{D},i} (t) = \int_{0}^{t} 2 \langle \Psi_i (t') | W | \Psi_i (t') \rangle \textrm{d}t'
\end{align}
where $i = \textrm{X}, \textrm{B}$.
The same quantities have a slightly different formula if the system is described by the density operator. To calculate the dissociation probability, we use the state $|\mathcal{D}\rangle$ introduced in Eq. \eqref{Eq:CAP-Lindblad},\textcolor{red}{}{\cite{94Sakurai}}
\begin{equation}
  \label{Eq:PD_Lindblad}
  P_{\textrm{D},|i,n\rangle} (t) = (\langle \mathcal{D} | \otimes \langle i,n|) \rho (t) (|i,n\rangle \otimes | \mathcal{D} \rangle)
\end{equation}
where $i = \textrm{X}, \textrm{B}$ and $n=0,1$.
The active or bound population can be expressed as \textcolor{red}{}{\cite{94Sakurai}}
\begin{equation}
  \label{Eq:PA_Lindblad}
  P_{\textrm{A},|i,n\rangle} (t) =  \langle i,n | \textrm{Tr}_\textrm{DVR} \left[ \rho (t) \right] | i,n \rangle - P_{\textrm{D},|i,n\rangle} (t)
\end{equation}
where $i = \textrm{X}, \textrm{B}$, $n=0,1$
and $\textrm{Tr}_\textrm{DVR}$ stands for a trace over the DVR basis of the vibrational DOF. 
The last term above is a technical byproduct, required as a consequence of the introduction of the helper state $|\mathcal{D}\rangle$.


In case of the non-Hermitian TDSE 
\textcolor{red}{}{we have no information about the state $|X,0\rangle$ since its wave function was absorbed}
by the imaginary term. The population represented by these wave packets can be easily calculated as the missing total population of the other states: $P_{|\textrm{X},0\rangle} (t) = 1 - \sum_{i=\textrm{X},\textrm{B}} \left( P_{\textrm{A},i} (t) + P_{\textrm{D},i} (t) \right)$. 
\textcolor{red}{}{To approximate (without the actual wave function)} the bound and dissociation probabilities of $|\textrm{X},0\rangle$, we use the ratio
\begin{align}
  \label{Eq:Splitting-ratio}
  \xi = \frac{P_{\textrm{D},|\textrm{X},0\rangle}}{P_{\textrm{A},|\textrm{X},0\rangle} + P_{\textrm{D},|\textrm{X},0\rangle}},
\end{align} 
\textcolor{red}{}{where the probabilities are obtained using the Lindblad method in which all (dynamically active) states are accessible.} The ratio above depends heavily on the central photon energy $\hbar \omega_\textrm{c}$ and slightly on the electric field strength $E$. 
The $\omega_\textrm{c}$ dependence was mapped directly by performing full propagations with different photon energy values, however the slight field strength dependence taken into account approximately: we determine an effective field strength $E_\textrm{eff}$ by the weighted average of $E \cos{\theta}$ with weights based on the angular distribution. Then, we use the ratio $\xi_{\omega_\textrm{c},E_\textrm{eff}}$ (obtained from a   \textcolor{red}{}{fixed-in-space} Lindblad calculation, with $\hbar\omega_c$ photon energy and $E=E_\textrm{eff}$ field strength) to approximate the population of $|\textrm{X},0\rangle$ as $P_{\textrm{A},|\textrm{X},0\rangle} \approx 1 - \xi \cdot P_{|\textrm{X},0\rangle}$ and dissociation probability as $P_{\textrm{D},|\textrm{X},0\rangle} \approx \xi \cdot P_{|\textrm{X},0\rangle}$.

\section{Results and Discussion}
We consider a H$_2$ molecule placed inside a cavity, focusing on its ground
electronic state X ($\textrm{X} \, ^{1}\Sigma_{g}^{+}$) and the 
excited electronic state B ($\textrm{B} \, ^{1}\Sigma_{u}^{+}$), 
both of which are singlet states. The potential energy surfaces 
$V_\textrm{X}(R)$ and $V_\textrm{B}(R)$, are coupled by the cavity photon.
As a result of cavity-molecule coupling, polaritonic states are formed.
We use a cavity central photon
energy of $\hbar\omega_{c}=7.4\,\mathrm{eV}$ because total dissociation occurs at maximum efficiency at this energy (as will be discussed later). 

We explore the scenario described as follows. Initially, the molecule is prepared in the
vibrational and rotational ground state of the electronic ground state.
\textcolor{red}{}{
Then, to initiate the dynamics ultrafast vertical excitation of the wave packet is presumed from $V_\textrm{X}(R)$ to $V_\textrm{B}(R)$.}
Applying the Lindblad master equation, one can
describe the temporal evolution of the nuclear wave packet dynamics \textcolor{red}{}{for the fixed-in-space model}. 
As the wave packet oscillates on the $|\textrm{B},0\rangle$
state, it reaches a region where the two diabatic states $|\textrm{B},0\rangle$ and $|\textrm{X},1\rangle$ 
are closest
(see Fig. \ref{fig:1}, panel A). At this point, part of the population is transferred
back to the $|\textrm{X},1\rangle$ photonic state. Due to the lossy
nature of the cavity, the population then decays back
{from $|\textrm{X},1\rangle$} 
to the ground
state $|\textrm{X},0\rangle$ where dissociation can take place.

\begin{figure}
\centering{}\includegraphics[width=0.49\textwidth]{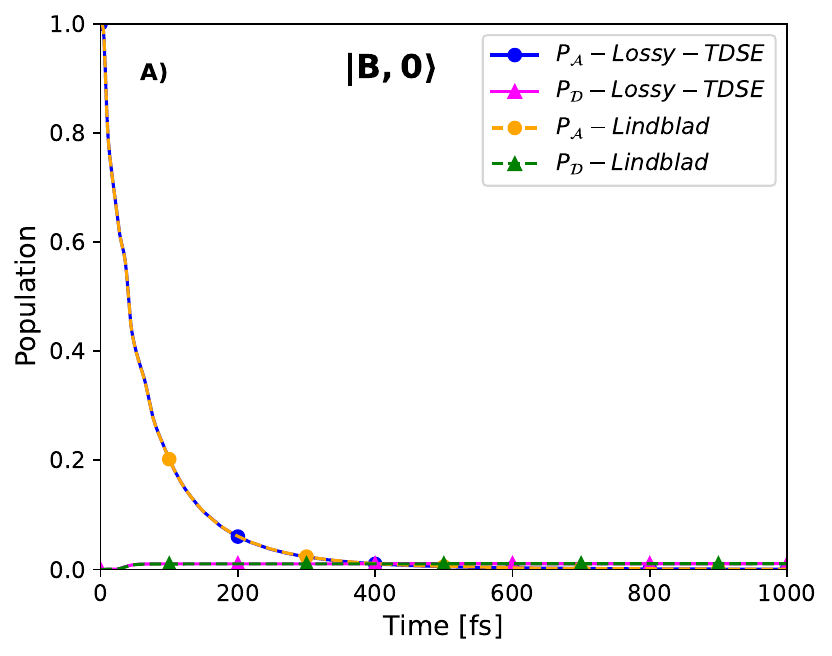}\\
\includegraphics[width=0.49\textwidth]{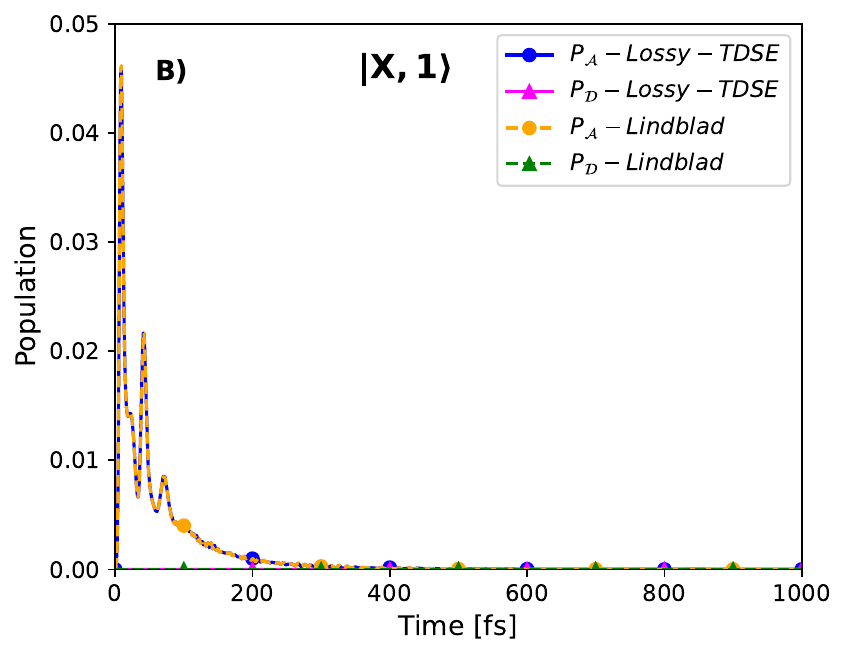}\\
\includegraphics[width=0.49\textwidth]{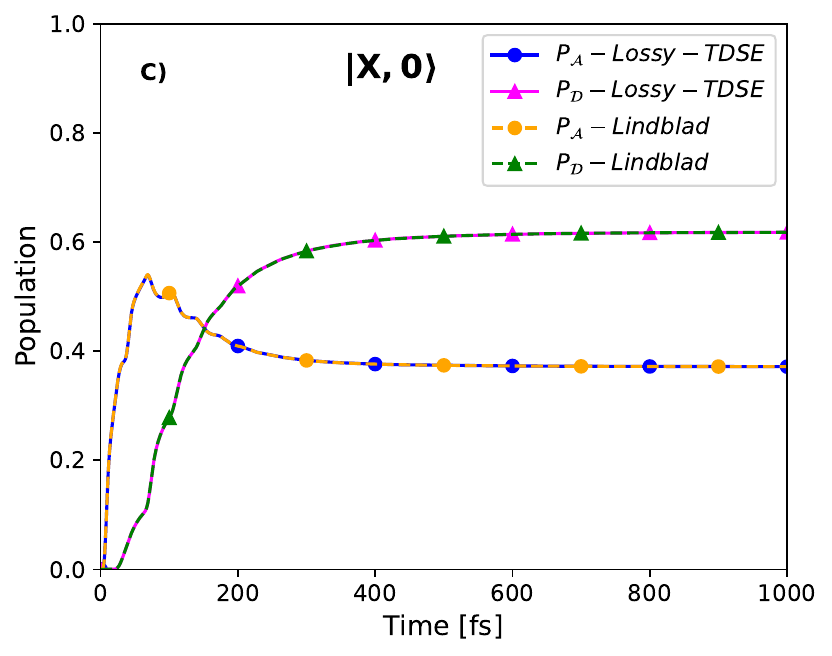}
\caption{ Populations ($P_\textrm{A}$) and dissociation probabilities ($P_\textrm{D}$)
of the three different diabatic states (panel A: $|\textrm{B},0\rangle$ state, panel B: $|\textrm{X},1\rangle$ state and panel C: $|\textrm{X},0\rangle$ state) using the Lindblad and the 
non-Hermitian lossy Schr\"odinger schemes    \textcolor{red}{}{with fixed-in-space molecules}. 
The parameters of the calculations are $\hbar\omega_\textrm{c}=7.4\,\textrm{eV}$ (cavity central photon energy), 
$E=70\,\,\mathrm{mV \slash a_0}$ \textcolor{red}{}{$\approx 2.57\cdot10^{-3}$ a.u.}
(electric field strength) and $\kappa=0.476\,\textrm{eV}$ (\textcolor{red}{}{$\tau\approx 1.38$ fs}). 
}
\label{fig:2} 
\end{figure}

Fig. \ref{fig:2} presents the population distributions and dissociation probabilities
for the three diabatic states. Notably, no dissociation occurs from
the $|\textrm{B},0\rangle$ state, and the population essentially vanishes
from this state until approximately $t=400 \, \textrm{fs}$. 
The population curve corresponding to the $|\textrm{X},1\rangle$ surface mirrors the shape of the
population in the $|\textrm{B},0\rangle$ state but is roughly 20 times
smaller in magnitude. However, from the intermediate $|\textrm{X},1\rangle$ state, the population
rapidly decays to the ground state $|\textrm{X},0\rangle$, from which
part of it dissociates and part of it remains bound. 
In the $|\textrm{X},0\rangle$
ground state, the population increases, reaching a maximum around $t=70 \, \textrm{fs}$,
and then decreases, stabilizing at a constant value by $t=400 \, \textrm{fs}$.
In contrast, the dissociation probability continues to rise slightly,
reaching a maximum at around $t=400 \, \textrm{fs}$, after which it
becomes essentially saturated. The time evolution is displayed
up to $t=1000 \, \textrm{fs}$ in Fig. \ref{fig:2}.

For comparison, we repeated the simulation using the non-Hermitian
TDSE approach, applying the same parameters as in the Lindblad scheme.
The non-Hermitian TDSE allows for the inclusion of a greater number
of nuclear DOFs (rotational and vibrational) in the
quantum-dynamical description, owing to its lower computational cost
compared to the Lindblad master equation. The results obtained from
the non-Hermitian TDSE are in perfect agreement with those derived
from the Lindblad equation (see Fig. \ref{fig:2}). In this \textcolor{red}{}{fixed-in-space} non-Hermitian TDSE numerical
description, 
the orientation of the molecular axis relative to the
cavity mode polarization direction remains fixed ($\theta=0$) throughout the simulation.
However, to describe appropriately the light-induced nonadiabatic
dynamics of the $\mathrm{H_{2}}$ molecule and to form a two-dimensional
branching space, it is necessary to extend the previous model by incorporating
the rotational DOF. 

\begin{figure}
\centering{}\includegraphics[width=0.33\textwidth]{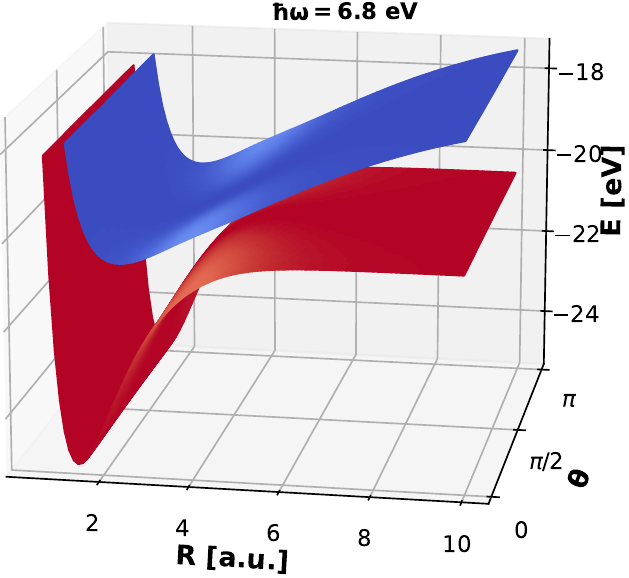}\hspace*{\fill} \includegraphics[width=0.33\textwidth]{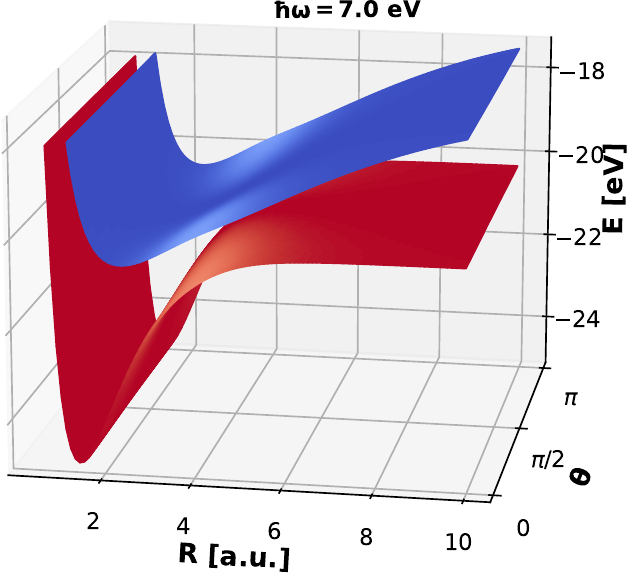}\hspace*{\fill} \includegraphics[width=0.33\textwidth]{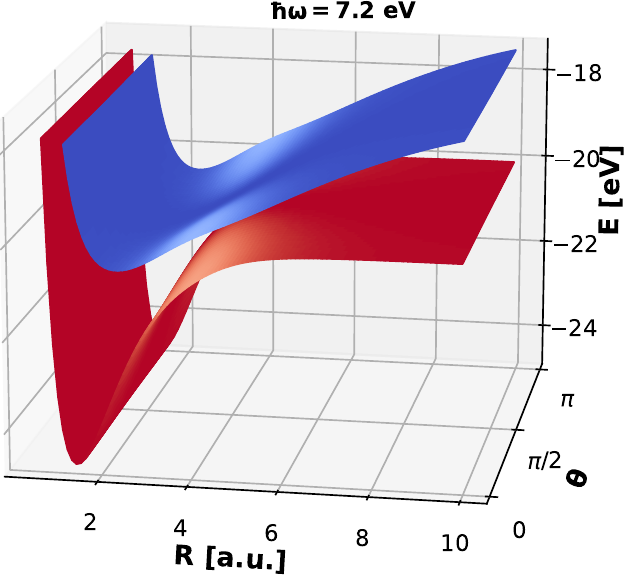}
\centering{}\includegraphics[width=0.33\textwidth]{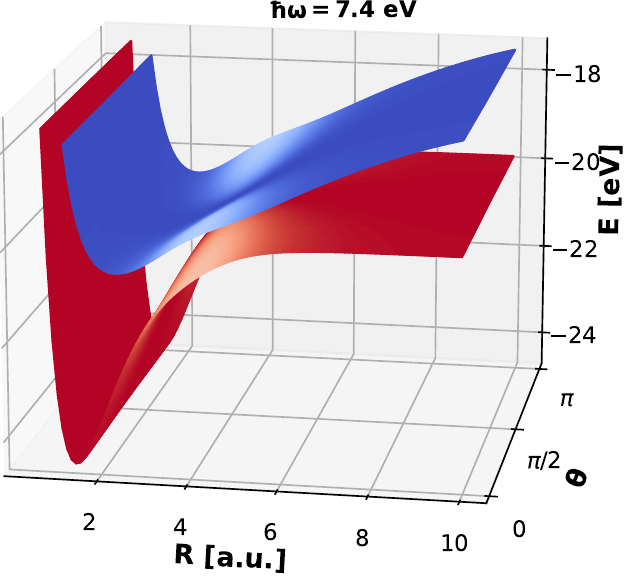}\hspace*{\fill} \includegraphics[width=0.33\textwidth]{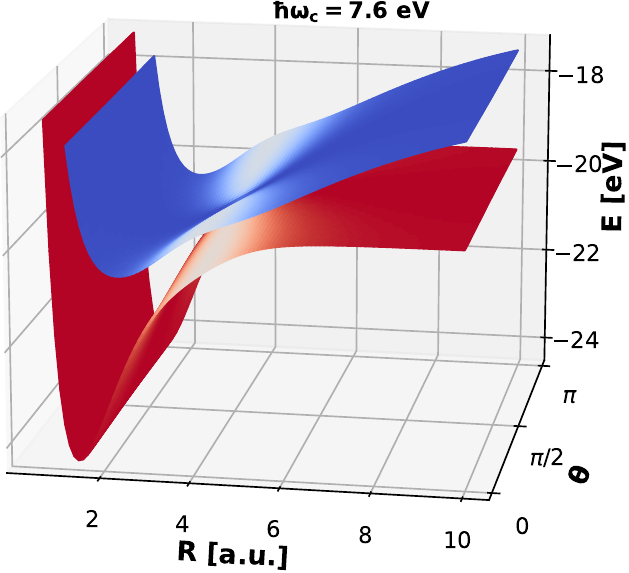}\hspace*{\fill} \includegraphics[width=0.33\textwidth]{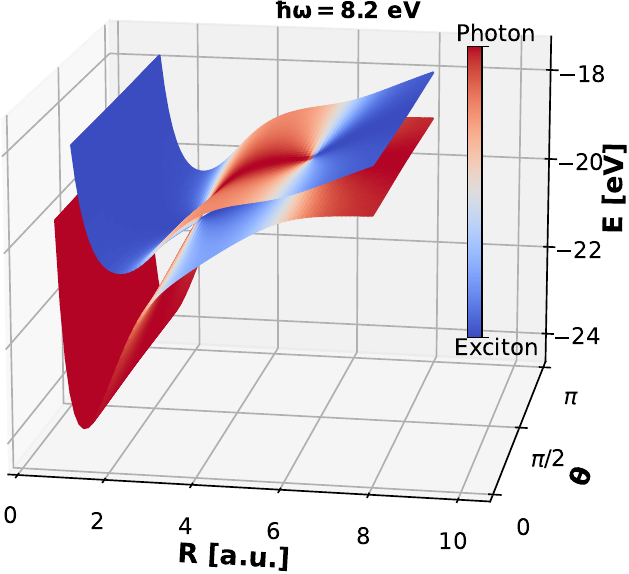}
\centering{}\includegraphics[width=0.33\textwidth]{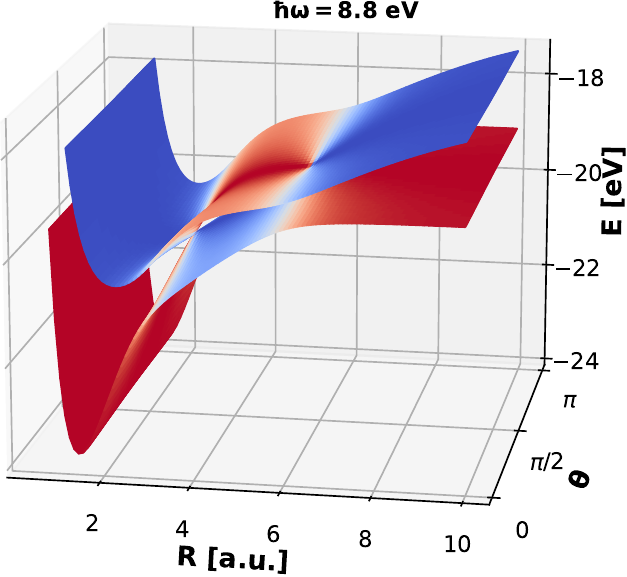}\hspace*{\fill} \includegraphics[width=0.33\textwidth]{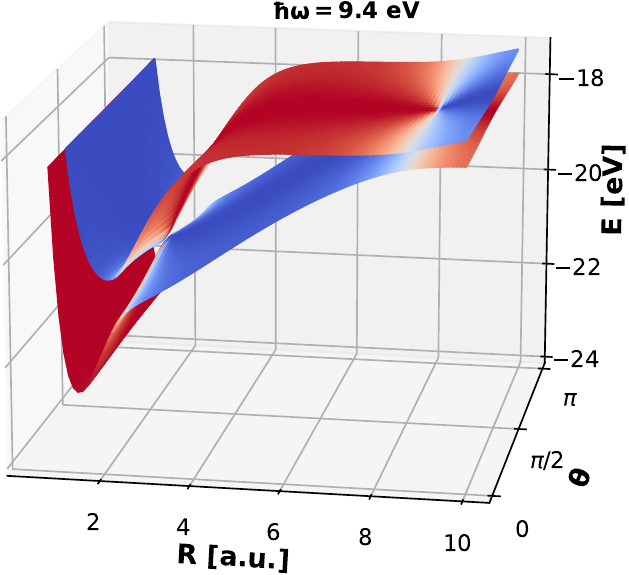}\hspace*{\fill} \includegraphics[width=0.33\textwidth]{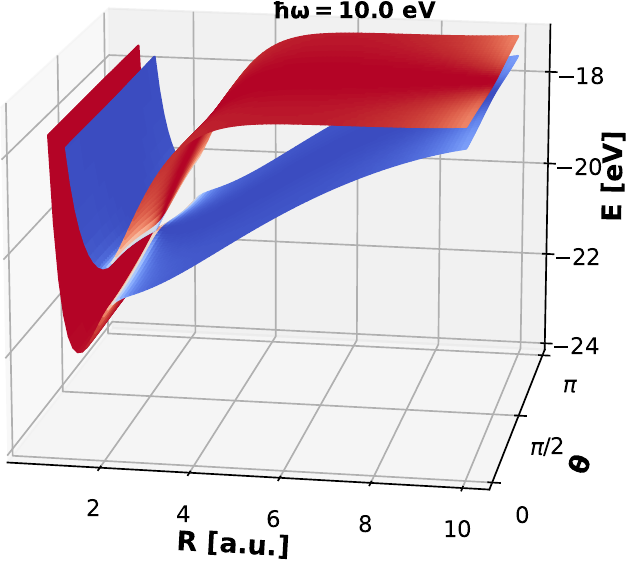}
\caption{Two-dimensional polaritonic potential energy surfaces along the vibrational
and rotational coordinates of the $\mathrm{H_{2}}$ molecule. Figures are shown for 
$\hbar\omega_\textrm{c}=7.6\,\textrm{eV}$ (cavity central photon energy) as well as for eight other energy values among the infinitely many that arise within the range of energy broadening. The $\hbar\omega_\textrm{c}=7.6\,\textrm{eV}$ value is the resonance energy as well, because the two potential energy curves are the closest to each other here.}
\label{fig:3}
\end{figure}

Subsequently, using the same parameter set that yielded excellent
agreement between the Lindblad and non-Hermitian TDSE schemes   \textcolor{red}{}{with fixed-in-space molecules}, we 
performed two-dimensional (2D) simulations. In this case, the rotational
DOF is treated as a dynamical variable during the numerical
calculations. This extended model enables the inclusion of light-induced
conical intersections (LICIs), thereby providing a more accurate treatment
of nonadiabatic quantum dynamics. In the 2D model, LICIs are
formed, as shown in panel C of Fig. \ref{fig:1}. In the lossy cavity scenario, the photon
energy undergoes broadening around the cavity central frequency, resulting
in the appearance of an infinite number of photon energies at both
lower and higher frequencies relative to the central frequency \cite{21ToFe,23SaRoKi}. 
At any given time, the nuclear dynamics are driven by this infinite set
of light-induced conical intersections, formed by couplings at the
corresponding photon energies. This situation is illustrated in Fig.
\ref{fig:3} where the lower and upper polariton  surfaces
are shown for selected cavity
frequencies in the range of $\hbar\omega_\textrm{c}=6.8\,\textrm{eV}-10\,\textrm{eV}$. 

\begin{figure}
\begin{centering}
\includegraphics[width=0.49\textwidth]{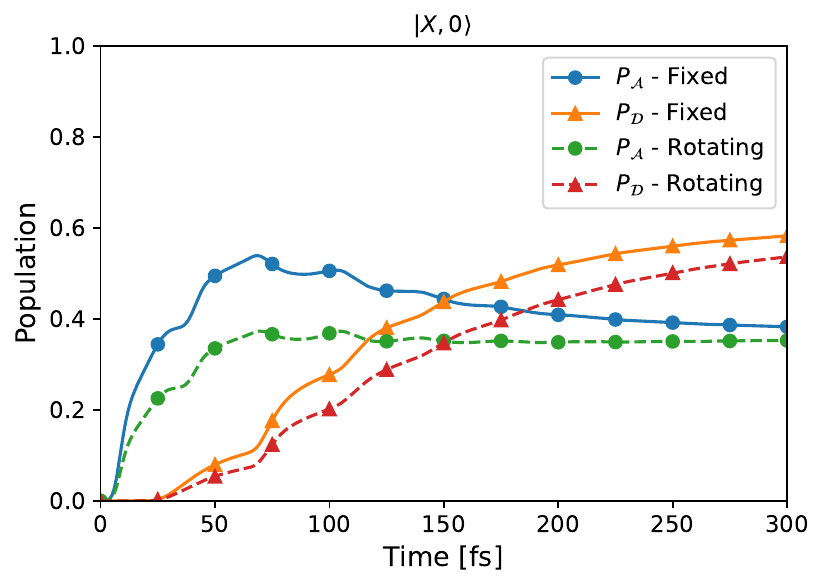}
\par\end{centering}
\caption{The active ($P_\textrm{A}$) and dissociated ($P_\textrm{D}$) populations
of the $|\textrm{X},0\rangle$ state as a function of time obtained for the case
of   \textcolor{red}{}{fixed-in-space} and rotating $\mathrm{H_{2}}$ molecules by using the non-Hermitian lossy TDSE. 
The parameters are the same as in Fig. \ref{fig:2}. 
}
\label{fig:4} 
\end{figure}

Let us now explore and discuss the   \textcolor{red}{}{fixed-in-space} and 2D 
nuclear dynamics in the non-Hermitian lossy TDSE framework. 
It is well established that conical intersections primarily
affect short-time dynamics, which is the reason why we restrict our analysis
to the $t=0-300 \, \textrm{fs}$ time interval. The results are
presented in Fig. \ref{fig:4}. At first glance, it is striking that the $|\textrm{X},0\rangle$
state is populated significantly faster in the   \textcolor{red}{}{fixed-in-space} model. Similarly,
molecular dissociation begins more rapidly from the $|\textrm{X},0\rangle$ state in the
\textcolor{red}{}{fixed-in-space} case. This is because the   \textcolor{red}{}{fixed-in-space} scheme facilitates more efficient
population transfer from the $|\textrm{B},0\rangle$ state to the $|\textrm{X},1\rangle$
state as the molecular axis remains parallel to the polarization direction
of the cavity mode throughout the dynamics.

In contrast, the 2D model can be interpreted as capturing the combined
effects of molecular rotation, light-induced conical intersections
(LICIs) and the varying magnitudes and directions of the respective
molecular transition dipole moments. As the nuclear wave packet oscillates
on the $|\textrm{B},0\rangle$ state, the molecule rotates.
When the wave
packet reaches the coupling region, part of the population is transferred
to the $|\textrm{X},1\rangle$ state. However, due to the rotation, the
angle $\mathrm{\theta}$ between the molecular transition dipole moment
and the photon polarization axis changes continuously, causing the
coupling strength to vary as well. As a result, less population is
transferred to the $|\textrm{X},1\rangle$ state compared to the   \textcolor{red}{}{fixed-in-space} case
where the photon polarization and the molecular axis remain aligned.

Moreover, when the wave packet reaches the positions of the LICIs,
nonadiabatic population transfer can take place between the upper
and lower polaritonic states, allowing part of the population to transfer
back from the $|\textrm{X},1\rangle$ state to the $|\textrm{B},0\rangle$
state. However, this effect is minor, as the population remains in
the $|\textrm{X},1\rangle$ state only for a very brief period. The combined
effect of these processes produces the 2D dynamical picture.
Additionally, due to the lossy nature of the cavity, photon energy
broadening results in the coupling of the $|\textrm{X},1\rangle$ and
$|\textrm{B},0\rangle$ surfaces at a range of frequencies, not just
the central frequency, further perturbing the dynamics. As is well
known, the higher the quality of the cavity, the more the photon loss
is concentrated around the cavity central frequency\cite{23SaRoKi}. 

\begin{figure}
\centering{}\includegraphics[width=0.49\textwidth]{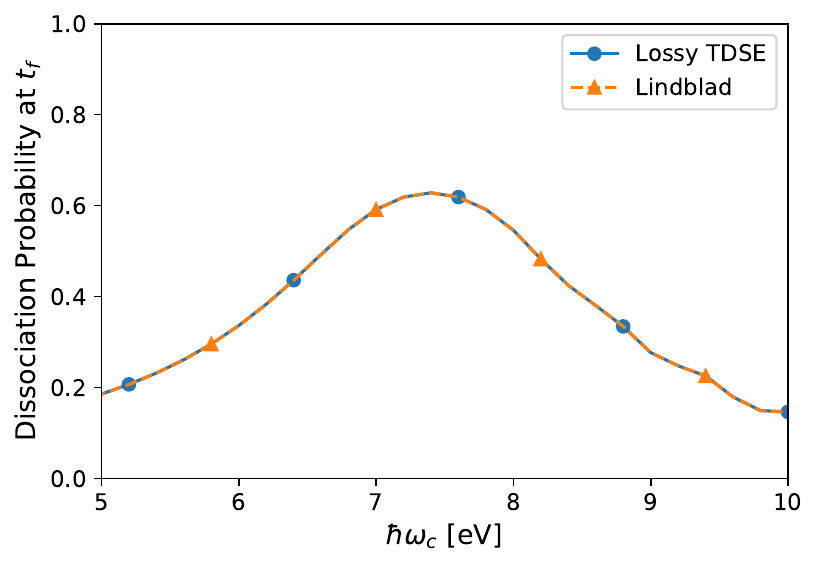}
\caption{ Dissociation probabilities ($P_\textrm{D}$) as a function
of cavity central photon energy at $t_\textrm{f} = 1000 \, \textrm{fs}$ obtained by the Lindblad and the  non-Hermitian
lossy Schr\"odinger schemes   \textcolor{red}{}{with fixed-in-space molecules}. 
The central photon energy ranges from $\hbar\omega_\textrm{c}=5\,\textrm{eV}$ to $\hbar\omega_\textrm{c}=10\,\textrm{eV}$ with steps of $\Delta \hbar\omega_\textrm{c}=0.2\,\textrm{eV}$. The electric field strength and loss rate were kept at $E=70\,\,\mathrm{mV \slash a_0}$ \textcolor{red}{}{$\approx 2.57\cdot10^{-3}$ a.u.} and $\kappa=0.476\,\textrm{eV}$ \textcolor{red}{}{$(\tau\approx 1.38$ fs)}, respectively.
}
\label{fig:5} 
\end{figure}

\begin{figure}
\centering{}\includegraphics[width=0.49\textwidth]{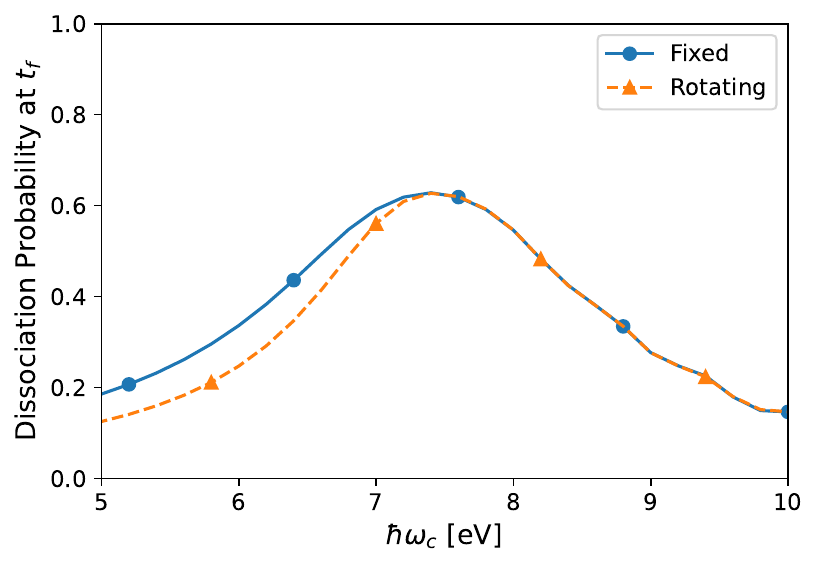}
\caption{ Dissociation probabilities ($P_\textrm{D}$) as a function
of cavity central photon energy at $t_\textrm{f} = 1000 \, \textrm{fs}$ obtained by the non-Hermitian lossy
Schr\"odinger equation
for the case of   \textcolor{red}{}{fixed-in-space} 
and rotating 
molecules. 
Parameters are the same as in Fig. \ref{fig:5}.
}
\label{fig:6} 
\end{figure}

To proceed, we set several values for the central photon energy
ranging from $\hbar\omega_\textrm{c}=5\,\textrm{eV}$ to $\hbar\omega_\textrm{c}=10\,\textrm{eV}$
and calculate the dissociation probabilities. 
As shown in Fig. \ref{fig:5},
the results from the Lindblad and the 
non-Hermitian lossy
Schr\"odinger equations   \textcolor{red}{}{with fixed-in-space molecules} are
nearly identical across all frequencies considered. 
This perfect agreement
allows us to conclude that, for a given range of frequencies and parameters,
the non-Hermitian Schr\"odinger equation provides an accurate description
of the nonadiabatic dissociation dynamics. Using the non-Hermitian Schr\"odinger equation, 
we then compared the dissociation probabilities obtained from the   \textcolor{red}{}{fixed-in-space} 
and 2D models. The results are shown in Fig. \ref{fig:6}.

It is evident that both the \textcolor{red}{}{fixed-in-space} 
and 2D models 
predict the highest dissociation probability
at central photon energy 
$\hbar\omega_\textrm{c}=7.4\,\textrm{eV}$. Interestingly,
the two 
\textcolor{red}{}{descriptions} yield different results for energies 
below $\hbar\omega_\textrm{c}=7.4\,\textrm{eV}$,
but they are quite similar for frequencies above. This disparity
arises because, for central energies
lower than $\hbar\omega_\textrm{c}=7.4\,\textrm{eV}$,
the photon energy broadening results in fewer coupling frequencies,
which cannot fully compensate for the combined effects of molecular
rotation and light-induced conical intersections (LICIs) compared
to the   \textcolor{red}{}{fixed-in-space} description. This holds true for the short-time dynamics
up to $t=1000\,\textrm{fs}$. In contrast, for central frequencies
higher than $\hbar\omega_\textrm{c}=7.4\,\textrm{eV}$ the $|\textrm{B},0\rangle$ and
$|\textrm{X},1\rangle$ states couple at many more frequencies, thereby
diminishing the combined effects of rotation and LICIs, which results
in dynamics nearly identical to the   \textcolor{red}{}{fixed-in-space} model.

\begin{figure}
\begin{centering}
\includegraphics[width=0.49\textwidth]{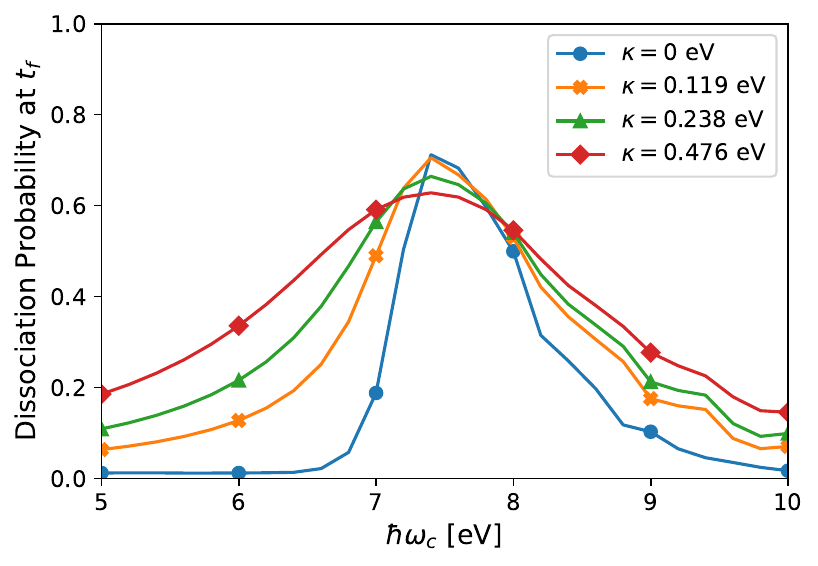} 
\par\end{centering}
\caption{ Dissociation probabilities ($P_\textrm{D}$) as a function
of cavity central photon energy at $t_\textrm{f} = 1000 \, \textrm{fs}$ obtained by the Lindblad master equation. \textcolor{red}{}{  \textcolor{red}{}{Fixed-in-space} molecules were applied for
the case of several decay rates: $\kappa=0$ eV ($\tau=\infty$), $\kappa=0.119$ eV ($\tau\approx 5.52$ fs), $\kappa=0.238$ eV ($\tau\approx 2.76$ fs) and $\kappa=0.476$ eV ($\tau\approx 1.38$ fs).  Other parameters are the same as in Fig. \ref{fig:5}.}
}
\label{fig:7} 
\end{figure}

Finally, we explored the effect of varying the cavity decay rate and
turning off the cavity leakage. In this scenario, only   \textcolor{red}{}{fixed-in-space} molecules
were considered. The results are presented in Fig. \ref{fig:7}.

As the cavity decay rate $\kappa$ decreases, the dissociation probability
curve becomes narrower, although its maximum value increases slightly.
The dissociation probability is at its narrowest when $\kappa=0$,
at which situation the dissociation probability reaches its maximum
value. This is because, with a finite cavity lifetime which is inversely
proportional to the decay rate, the photon energy broadens, causing
photons with higher energies than the central frequency to appear.
As a result, for a given central coupling frequency, dissociation can also
occur at energies below $\hbar\omega=7.4\,\textrm{eV}$. The higher the
cavity decay rate, the shorter the lifetime, and thus the broader
the photon energy distribution. At $\kappa=0$, however,
dissociation almost completely disappears at frequencies lower than
the central frequency $\hbar\omega_\textrm{c}=7.4\,\textrm{eV}$, as the system
cannot couple at those energies due to the absence of losses (infinite
resonator lifetime). Consequently, the population cannot decay to
the ground state $|\textrm{X},0\rangle$, but rather remains in the $|\textrm{X},1\rangle$ state, which is the photonic copy of the ground state,
and dissociates from
there. Due to the infinite cavity lifetime, the energy broadening
vanishes entirely, and the resonance condition is perfectly satisfied,
leading to the highest dissociation rate from the $|\textrm{X},1\rangle$
state.

For higher central frequencies, dissociation occurs at different magnitudes.
In these cases, the $|\textrm{X},1\rangle$ and $|\textrm{B},0\rangle$
states always intersect, and the dynamics are determined by the coupling
frequencies of the photon energy broadening in the lossy cavity. In
the central frequency range $\hbar\omega_\textrm{c}=7.4\,\textrm{eV}-9\,\textrm{eV}$,
the lossless and lossy cases differ significantly, with the lossless dissociation
decaying earlier. In this range, the specific value of $\kappa$ does
not play a significant role. However, at higher frequencies, the dissociation
rate increases with higher losses and shorter photon lifetimes.

It is noteworthy that dissociation still occurs even when $\mathrm{\kappa=0}$.
This observation is consistent with the fact that the long-wavelength
Floquet picture and the cavity description in the first excited state
manifold are identical when the central frequency is the resonance
frequency (i.e., when the two states are closest to each other). However,
when higher excited states are included, the two descriptions are
different\cite{24AvWaYa,24FaCsHa_2}.
\section{Conclusions}
We have explored the combined effects of molecular rotations 
and light-induced conical intersections (LICIs) on nonadiabatic quantum
dynamics within a lossy cavity. Due to the bandwidth surrounding the
central cavity frequency $\omega_\textrm{c}$ multiple light-induced conical
intersections can form, influencing the dynamics in ways that differ
from the \textcolor{red}{}{fixed-in-space} description. The impact of LICIs depends on the relationship
between the central cavity frequency and the resonance frequency between
the $|\textrm{X},1\rangle$ and $|\textrm{B},0\rangle$ states. When the
cavity central frequency is below $\hbar\omega_{c}=7.4\,\textrm{eV}$, the dissociation
probability is reduced, while for frequencies above resonance, both the \textcolor{red}{}{fixed-in-space} and two-dimensional 
descriptions yield nearly identical results. Notably, due to energy broadening induced by cavity losses, dissociation also occurs at central frequencies well below the cavity resonance frequency.

Changes in the cavity lifetime---whether it decreases or increases---directly
affect the loss rate. Larger loss rates result in greater photon
energy broadening. This, in turn, alters the dissociation probability,
either enhancing or diminishing it. In the extreme case where cavity
losses vanish, the cavity lifetime becomes infinite. In this limit,
with finite photon energy and within the framework of the first excited-state manifold, 
the behavior of the system is equivalent to the well-known
Floquet description used in laser physics, which applies to continuous
wave limits.

\begin{acknowledgement}
The authors are indebted to NKFIH for funding (Grant No. K146096).
This paper was supported by the J\'anos
Bolyai Research Scholarship of the Hungarian Academy of Sciences and
the University of Debrecen Program for Scientific Publication. The authors thank Johannes Feist for valuable contributions. We appreciate fruitful discussions with Lorenz S. Cederbaum and St\'ephane Gu\'erin.
\end{acknowledgement}


\bibliography{H2_cavity}

\end{document}